\documentclass[conference]{IEEEtran}
\IEEEoverridecommandlockouts
\usepackage{listings}
\usepackage{makeidx}
\usepackage{algorithm,algpseudocode}
\usepackage{graphicx}
\usepackage{float}
\usepackage{cite}
\usepackage{amsmath}
\usepackage{rotating,todonotes, xspace}
\usepackage{multirow}
\usepackage{hyperref}
\usepackage{xspace}
\usepackage{paralist}
\usepackage{booktabs}
\usepackage{balance}
\usepackage{cleveref}
\usepackage{todonotes}
\usepackage{caption}
\usepackage{subcaption}                    % Source code
\usepackage{algorithm}                      % Pseudo Code
\usepackage{algpseudocode}
\usepackage{varwidth}

\makeatletter
\let\OldStatex\Statex
\renewcommand{\Statex}[1][3]{%
  \setlength\@tempdima{\algorithmicindent}%
  \OldStatex\hskip\dimexpr#1\@tempdima\relax}
\makeatother

\newcommand{\alg}{\textsc{WfChefRecipe}\xspace}
\newcommand{\alggenerate}{\textsc{WfChefGenerate}\xspace}
\newcommand{\algreplicate}{\textsc{ReplicatePOs}\xspace}

\newcommand{\tool}{WfChef\xspace}
\newcommand{\generator}{WorkflowGenerator\xspace}
\newcommand{\workflowhub}{WorkflowHub\xspace}

\newtheorem{definition}{Definition}[subsection]

\begin{document}

\title{WfChef: Automated Generation of Accurate Scientific Workflow Generators}
%\title{WfChef: A Synthetic Scientific Workflow Generator Generator}

\author{
  \IEEEauthorblockN{
    Tain\~a Coleman\IEEEauthorrefmark{1},
    Henri Casanova\IEEEauthorrefmark{2},
    Rafael Ferreira da Silva\IEEEauthorrefmark{1}
  }
  \IEEEauthorblockA{
    \IEEEauthorrefmark{1}Information Sciences Institute, University of Southern California, Marina Del Rey, CA, USA \\
    \IEEEauthorrefmark{2}Information and Computer Sciences, University of Hawaii, Honolulu, HI, USA \\
    \{tcoleman,rafsilva\}@isi.edu, henric@hawaii.edu \\
  }
}

\maketitle
\thispagestyle{empty}
\pagestyle{empty}

%%%%%%%%%%%%%%%%%%%%%%%%%%%%%%%%%%%%%%%%%%%%%%%%%%%%%%%%%%%%%%%%
\begin{abstract}
% \HC{Drafted the abstract, see what you think}
Scientific workflow applications have become mainstream and their automated
and efficient execution on large-scale compute platforms is the object of
extensive research and development. For these efforts to be successful, a
solid experimental methodology is needed to evaluate workflow algorithms
and systems. A foundation for this methodology is the availability of
realistic workflow instances.  Dozens of workflow instances for a few
scientific applications are available in public repositories.  While these
are invaluable, they are limited: workflow instances are not available for
all application scales of interest. To address this limitation, previous
work has developed generators of synthetic, but representative, workflow
instances of arbitrary scales.  These generators are popular, but
implementing them is a manual, labor-intensive process that requires expert
application knowledge.  As a result, these generators only target a handful
of applications, even though hundreds of applications use workflows in
production.

In this work, we present \emph{\tool}, a framework that fully automates the
process of constructing a synthetic workflow generator for any
scientific application. Based on an input set of workflow instances, \tool
automatically produces a synthetic workflow generator.  We define and
evaluate several metrics for quantifying the realism of the generated
workflows.  Using these metrics, we compare the realism of the
workflows generated by \tool generators to that of the workflows
generated by the previously available, hand-crafted generators. We find
that the \tool generators not only require zero development effort (because
it is automatically produced),  but also generate workflows that are more
realistic than those generated by hand-crafted generators.

\end{abstract}

\begin{IEEEkeywords}
Scientific workflows, synthetic workflow generation, workflow management systems
\end{IEEEkeywords}

%%%%%%%%%%%%%%%%%%%%%%%%%%%%%%%%%%%%%%%%%%%%%%%%%%%%%%%%%%%%%%%%

% sections
\section{Introduction}
\label{sec:introduction}

% \HC{As usual the word "trace" is confusing, and weirdly conflated with the word "workflow", "configuration", "graph", "execution log"..... always hated that word :)}

Many computationally intensive scientific applications have been framed as
\emph{scientific workflows} that execute on various compute platforms and
platform scales~\cite{liew2016scientific}.  Scientific workflows are
typically described as Directed Acyclic Graphs (DAGs) in which vertices
represent tasks and edges represent dependencies between tasks, as defined
by application-specific semantics.
%As workflows continue to be adopted by
%scientific projects and user communities, they are becoming more complex.
%Today's production workflows can be composed of millions of individual
%tasks that execute for milliseconds to hours, and that can be
%single-threaded programs, multi-threaded programs, tightly coupled parallel
%programs (e.g., MPI programs), or loosely coupled parallel programs (e.g.,
%MapReduce jobs), all within a single workflow~\cite{ferreiradasilva-fgcs-2017}.
The automated execution of
workflows on these platforms have been the object of extensive research
and development, as seen in the number of proposed workflow resource
management and scheduling approaches\footnote{The IEEE Xplore digital
database includes 118 articles with both the words ``Workflow" and
``Scheduling" in their title for 2020 alone.}, and the number of developed
workflow systems (a self-titled ``incomplete" list\footnote{\url{https://s.apache.org/existing-workflow-systems}}
points to 290+ distinct systems, although many of them are no longer in use).
Thus, in spite of workflows and workflow systems being used in production
daily, workflow computing is an extremely active research and development
area, with many remaining challenges~\cite{ferreiradasilva-fgcs-2017,
ferreiradasilva2021wcs}.

Addressing these challenges requires a solid experimental methodology for
evaluating and benchmarking workflow algorithms and systems.  A fundamental
component of this methodology is the availability of sets of representative
workflow instances.  One approach is to infer workflow structures from
real-world execution logs.  We have ourselves followed this approach in
previous work~\cite{ferreiradasilva-escience-2014,
ferreiradasilva2020works}, resulting in a repository that provides $\sim$20
workflow instances for each of a handful of scientific applications.
These instances have been used by researchers, often for driving
simulation experiments designed to evaluate scheduling and resource
management algorithms.

Real workflow instances are by definition representative of real applications, but
they cover only a limited number of scenarios.  To overcome this
limitation, in previous work we have developed tools for generating
synthetic workflows by extrapolating the patterns seen in real
workflow instances.  The work
in~\cite{ferreiradasilva-escience-2014} presented a synthetic workflow
generator for four workflow applications, which has been used extensively by
researchers\footnote{To date, 300+ bibliographical references to the
research article and/or the software repository's URL.}.  The method for
generating the synthetic workflows was ad-hoc and based on expert knowledge
and manual inspection of real workflow instances. Our more recent generator
in~\cite{ferreiradasilva2020works} improves on the previous generator by
using information derived from statistical analysis of execution logs.
It was shown to generate more realistic workflows than the earlier
generator, and in particular to preserve key workflow features when
generating workflows at different scales~\cite{ferreiradasilva2020works}.
The main drawback of these two generators is that implementing the workflow
generation procedure  is labor-intensive.  Generators are manually crafted
for each application, which not only requires significant development
effort (several hundreds of lines of code) but also, and more importantly,
expert knowledge about the scientific application semantics that define
workflow structures.  As a result, this
approach is not scalable if synthetic workflow instances are to be generated
for a large number of scientific applications.

% \HC{Is the "(almost)" below needed?}

In this work, we present \textbf{\emph{\tool}}, a framework
that \emph{automates} the process of constructing a synthetic workflow
generator for any given workflow application. \tool takes as input a set
of real workflow instances from an application, and
outputs the code of a synthetic workflow generator for that application.
\tool analyzes the real workflow graphs in order to identify subgraphs that
represent fundamental task dependency patterns.  Based on the
identified subgraphs and on measured task type frequencies in the real
workflows, \tool outputs a generator that can generate realistic synthetic
workflow instances with an arbitrary numbers of tasks. In this work, we
evaluate the realism of the synthetic workflows generated by our approach,
both in terms of workflow structure and execution behavior.
Specifically, this work makes the following  contributions:

\begin{compactenum}

  \item We describe the overall architecture of \tool and the algorithms it
        uses to analyze real workflow instances and produce a workflow
        generator;

  \item We quantify the realism of the generated workflows when compared
        to real workflow instances, in terms of
        abstract graph similarity metrics and of the realism of
        simulated workflow executions;

  \item We compare the realism of the generated workflows to that of the
        workflows generated by the original workflow generator
        in~\cite{ferreiradasilva-escience-2014} and by the more recent generator
        in~\cite{ferreiradasilva2020works};

   \item Our key finding is that the generators automatically produced by
    \tool lead to equivalent or improved (often vastly) results when
    compared to the previously available, manually implemented, workflow
    generators.

\end{compactenum}

%\HC{Now that we have error bars for results, perhaps we need to tone down the last bullet above}

This paper is organized as follows.
Section~\ref{sec:relatedwork} discusses related work.
Section~\ref{sec:problem} defines our target problem.
Section~\ref{sec:approach} describes \tool.
Section~\ref{sec:evaluation} presents experimental evaluation results.
Finally, Section~\ref{sec:conclusion} concludes with a summary of results
and perspectives on future work.

\section{Related Work}
\label{sec:relatedwork}

Scientific workflow configurations, both inferred from real-world executions
and synthetically generated, have been used extensively in the workflow
research and development community, in particular for evaluating resource management and
scheduling approaches.  As scientific workflows are typically represented
as Directed Acyclic Graphs (DAGs), several tools have been developed to generate random DAGs, based
on specified ranges for various parameters~\cite{daggen,
amer2012evaluating, amalarethinam2011dagen, amalarethinam2012dagitizer}.
For instance, DAGGEN~\cite{daggen} and SDAG~\cite{amer2012evaluating}
generate random DAGs based on ranges of parameters such as the number
of tasks, the width, the edge density, the maximum number of levels
that can be spanned by an edge, the data-to-computation ratio, etc.
Similarly, DAGEN~\cite{amalarethinam2011dagen} generates random DAGS, but
does so for parallel programs in which the task computation and
communication payloads are modeled according to actual parallel programs.
DAGITIZER~\cite{amalarethinam2012dagitizer} is an extension of DAGEN for
grid workflows where all parameters are randomly generated. Although these
generators can produce a very diverse set of DAGs, they may not
resemble those of actual scientific workflows as they do not capture patterns
defined by application-specific semantics.

An alternative to random generation is to generate DAGs based on the
structure of real workflows for particular scientific applications.
In~\cite{van2003workflow}, over forty workflow patterns are identified for
addressing business process requirements (e.g., sequence, parallelism,
choice, synchronization, etc.). Although several of these patterns can be
mapped to some extent to structures that occur in scientific workflows~\cite{yildiz2009towards}, they do not fully capture these
structures.  In particular, they do not necessarily respect the ratios of
different types of particular workflow tasks in these structures. This 
is important because
a workflow structure is not only defined by a set of vertices and edges,
but also by the task type (e.g., an executable name) of each vertex.  The
work in~\cite{garijo2014common} focuses on identifying workflow ``motifs"
based on observing the data created and used by workflow tasks so as to
reverse engineer workflow structures. These motifs capture workflow
(sub-)structures, and can thus be used for automated workflow generation.
Unfortunately, identifying these motifs is an arduous manual
process~\cite{garijo2014common}.  In our previous
work~\cite{ferreiradasilva-escience-2014}, we developed a tool for
generating synthetic workflow configurations based on real-world workflow
instances.  Although the overall structure of generated workflows was
reasonably realistic, we found that workflow execution (simulated) behavior
was not (see~\cite{ferreiradasilva2020works} and also results in
Section~\ref{sec:eval-comparison}).  In~\cite{ferreiradasilva2020works}, we developed
an enhanced version of that earlier generator in which task
computational loads are more accurately captured by using statistical
methods. As a result, the generated synthetic workflows are more realistic
when compared to real-world workflows.
While the task computational load characterization is
automated, the DAG-generation procedure is labor-intensive because generators
are manually crafted
and rely on expert knowledge of the workflow application.

To the best of our knowledge, this is the first work that attempts a
completely automated synthetic workflow generation approach (automated
analysis of real workflow instances to drive the automated generation of
synthetic workflows).  Our approach makes it straightforward to generate
synthetic workflows for arbitrary scales that are representative of real
workflow instances for any workflow application.  These synthetic workflows
are key for supporting the development and evaluation of workflow
algorithms. Also, they can provide a fundamental building block for the
automatic generation of workflow application
skeletons~\cite{katz2016application}, which can then be used to benchmark
workflow systems.

% \HC{Discuss related work for workflow/graph generation}

\section{Problem Statement}
\label{sec:problem}

% \HC{Pick one: vertex of node. node is fine. task is ok.}

% \HC{Say something below about the fact that the non-parallel task
% assumption is still representative of many current apps?}
%\HC{I formally defined $v_i^j$ and $e_i^{k,k}$ below, we'll see if we use them in upcoming sections...}
%\HC{I opted to "v" for vertex, and not using node at all, since it can be confused with compute node}

Consider a scientific application for which a list of real workflow
instances, $W$, is available. Each workflow
$w$ in $W$ is a DAG, where the vertices represent workflow tasks and the edges
represent task dependencies.  In this work, we only consider workflows that
comprise tasks that execute on a single compute node -- i.e., tasks are not
parallel jobs (which is the case for a large number of scientific workflow
applications~\cite{ferreiradasilva-fgcs-2017, liew2016scientific,
liu2015survey, malawski2017serverless}).  More formally, $w = (V, E)$,
where $V$ is a set of vertices and $E$ is a set of directed
edges.  We use the notation $|w|$ to denote the number of vertices in $w$ (i.e.,
$|w| = |V|$).
We assume that each workflow has a single
entry vertex and a single exit vertex (for workflows that do not we
simply add dummy entry/exit vertices with necessary edges to all actual
entry/exist vertices).
%Without loss of generality, we assume that the workflows
%in $W$ are sorted by non-decreasing $|w|$.
Finally, a \emph{type} is associated to each
vertex $v$, denoted as $type(v)$. This type denotes the particular computation that the
corresponding workflow task must perform. In this work, we consider
workflows in which every task corresponds to an invocation of a particular
executable, and we simply define a vertex's type as the name of that
executable. Several tasks in the same workflow can thus have the same type.

\medskip
\noindent{\bf Problem Statement --}
Given $W$, the objective is to produce the code for a workflow generator that
generates realistic synthetic workflow instances.
This workflow generator takes as input an integer, $n \geq \min_{w \in
W}(|w|)$.  It outputs a workflow $w'$ with $n' \geq n$ vertices that is as
realistic as possible. $n'$ may not be equal to $n$, because real worfklows
for most scientific applications cannot be feasibly instantiated for
arbitray numbers of tasks. Our approach guarantees that $n'$ is the
smallest feasible number of tasks that is greater than $n$.

We use several metrics to quantify the realism of the generated
workflow. Consider a workflow generated with the workflow generator,
$w'$, and a real workflow instance with the same number of vertices, $w$.
The realism of workflow $w'$ can be quantified based on DAG similarity
metrics that perform vertex-to-vertex and edge-to-edge
comparisons (see Section~\ref{sec:eval-structure}).  The realism can also be quantified
based on similarity metrics computed between the logs of (simulated) executions
of workflows $w$ and $w'$ on a given compute platform (see
Section~\ref{sec:eval-comparison}).

\section{The \tool Approach}
\label{sec:approach}

In this section, we describe our approach, \tool. In
Section~\ref{sec:structures}, we define particular sub-DAGs in a set of
workflow instances. Algorithms to detect these sub-DAGs and use them for synthetic  workflow
generation are described in Section~\ref{sec:algorithms}.
Finally, in Section~\ref{sec:implementation} we briefly describe our
implementation of \tool.

\subsection{Pattern Occurrences}
\label{sec:structures}

The basis for our approach is the identification of particular
sub-DAGS in workflow instances for a particular application.
Let us first define the concept of a \emph{type hash}:

% \begin{definition}[\textbf{Type hash}]
%     \label{def:th}
%     Given a workflow vertex $v$, we recursively define its
%     \emph{top-down hash}, $TD(v)$, as the following string.
%     Consider the set of unique top-down hashes of
%     $v$'s successors. $TD(v)$ is the
%     lexicographically sorted concatenation of these hashes. $TD(v)$
%     is the empty string if $v$ is an entry vertex. We define
%     $v$'s \emph{bottom-up hash}, $BU(v)$, similarly, but considering
%     predecessors instead of successors, so that $BU(v)$ is the
%     empty string if $v$ is an exit vertex.  Finally, we define $v$'s
%     \emph{type hash}, $TH(v)$, as the concatenation of $type(v)$,
%     $TD(v)$, and $BU(v)$.
% \end{definition}

\begin{definition}[\textbf{Type hash}]
    \label{def:th}
    Given a workflow vertex $v$, we define its
    \emph{top-down hash}, $TD(v)$, recursively as the following string.
    Consider the lexicographically sorted list of the unique top-down hashes of
    $v$'s successors. $TD(v)$ is the
    concatenation of these top-down hashes and of $type(v)$. We define
    $v$'s \emph{bottom-up hash}, $BU(v)$, similarly, but considering
    predecessors instead of successors. Finally, we define $v$'s
    \emph{type hash}, $TH(v)$, as the concatenation of $TD(v)$ and $BU(v)$.
\end{definition}

\begin{figure}[h]
    \centering
    \includegraphics[width=1.0\linewidth]{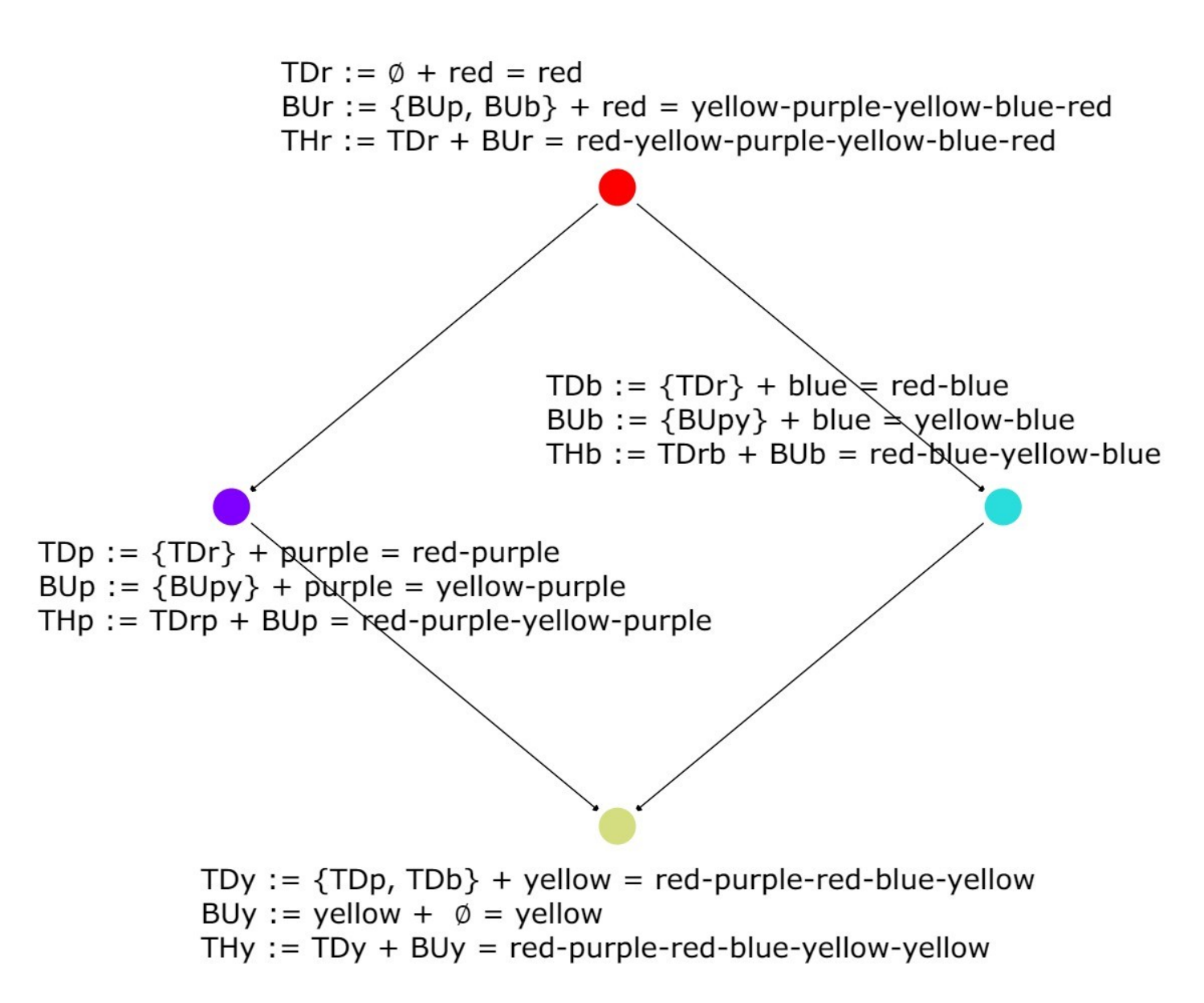}
    \caption{Example workflow with $TD$, $TU$, and $TH$ strings shown for
    each vertex. Vertex types are simply their colors (``red", ``purple", ``blue", or ``yellow"). $\emptyset$ denotes the empty string, and $+$ denotes the
    string concatenation operator.}
    \label{fig:TH}
\end{figure}

% \HC{Font sizes in Fig~\ref{fig:TH} could perhaps be increased}

Figure~\ref{fig:TH} shows an example for a simple 4-task diamond workflow,
where $TD$, $TU$, and $TH$ strings are shown for each vertex.
The type hash of each vertex in a workflow encodes information regarding
the vertex's role in the structures and sub-structures of the workflow.
From now on, we assume that each vertex is annotated with its type hash.
Given a workflow $w$, we define the type hash of $w$, denoted as $TH(w)$,
as the set of unique type hashes of $w$'s vertices. $TH(w)$ can be computed
in $O(|w|^2 \log(|w|))$.

The basis of our approach is the observation that, given a workflow,
sub-DAGs of it that have the same type hash are representative of the same
application-specific pattern (i.e., groups of vertices of certain types
with certain dependency structures, but not necessarily the same size).  We
formalize the concept of a \emph{pattern occurrence} as follows:

\begin{definition}[\textbf{Pattern Occurrence (PO)}]
    Given $W$, a set of workflow instances for an application,
    a \emph{pattern occurrence} is a DAG $po$ such that:
    \begin{compactitem}
      \item $po$ is a sub-DAG of at least one workflow in $W$;
      \item There exists at least one workflow in $W$ with two
            sub-DAGs $g'$ and $g''$ such that:
            \begin{compactitem}
              \item $g'$ and $g''$ are disjoint;
              \item $TH(g') = TH(g'') = TH(po)$;
	            \item Any two entry, resp. exit, vertices in $g'$ and
		                $g''$ that have the same type hash have the exact same
		                parents, resp. children.
            \end{compactitem}
    \end{compactitem}
\end{definition}

\begin{figure}[!t]
	\centering
	\includegraphics[trim={3cm 3cm 3cm 3cm}, clip,angle=270, width=\linewidth]{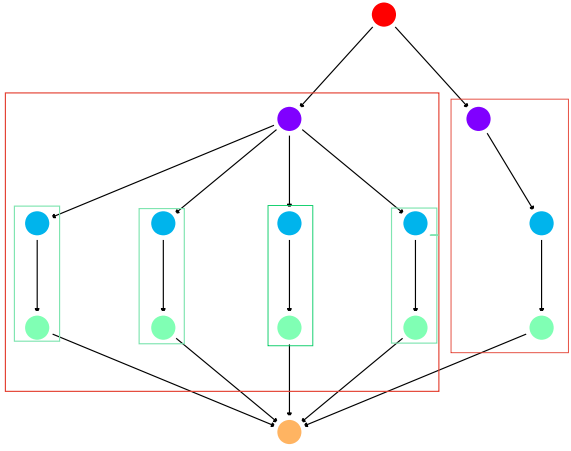}
    \caption{Example workflow with 6 POs, shown in rectangular boxes. Boxes with the same color indicate POs with identical type hashes.}
    \label{fig:example}
\end{figure}

Figure~\ref{fig:example} shows an example workflow, where vertex types are once again
indicated by colors. Based on the above definitions, this workflow contains
6 POs, each shown within a rectangular box.  The two POs in the red boxes
have the same type hash, and we say that they correspond to the same pattern.
But note that although they correspond to the same pattern, they do not
have the same number of vertices.  POs can occur within POs, as is the case
for the POs in the green boxes in this example.  Note that a sub-DAG of
the rightmost POs (the three-task PO in the red box) has the same
type hash as the POs in the green boxes. In fact, it is identical to those
POs (i.e., a blue vertex followed by a green vertex). But
this subgraph is not a PO because it does not have a common
ancestor with any of the other POs with similar type hashes.

%\begin{figure}[!t]
%	\centering
%	\includegraphics[trim={3cm 3cm 3cm 3cm}, clip, width=.8\linewidth]{figs/branch.pdf}
%    \caption{Example of a branch detected as a structure. The subgraph highlighted in bold is the structure.}
%    \label{fig:branch_structure}
%\end{figure}

\subsection{Algorithms}
\label{sec:algorithms}

\tool consists of two main algorithms, \alg and \alggenerate. The former is
invoked only once and takes as input a set of workflow instances for a
particular application, $W$, and outputs a ``recipe", i.e., a data
structure that encodes relevant information extracted from the workflow
instances.  The latter is invoked each time a synthetic workflow instance
needs to be generated. It takes as input a recipe and a desired number of
vertices (as well as a seeded pseudo-random number generator), and outputs
a synthetic workflow instance.  Both these algorithms have
polynomial complexity and implement several heuristics, as described
hereafter.

\algnewcommand\algorithmicforeach{\textbf{for each}}
\algdef{S}[FOR]{ForEach}[1]{\algorithmicforeach\ #1\ \algorithmicdo}

\begin{algorithm}[!ht]
  \begin{algorithmic}[1]
      \Function{\alg}{$W$}
      \State $POs \gets \{\}$  \Comment dictionary of POs
      \ForEach{$w \in W$}
          \State $POs[w] \gets []$ \Comment list of POs in $w$
          \ForEach{unvisited vertex $v$ in $w$}
              \State mark $v$ as visited
              \State $v' = $ an unvisited vertex s.t. $TH(v') = TH(v)$
              \State {\bf if} {$v'$ is not found} {\bf then continue}
              \State mark $v'$ as visited
              \State A = \textsc{ClosestCommonAncestors}($v$,$v'$)
              \State D = \textsc{ClosestCommonDescendants}($v$,$v'$)
              \State {\bf if} $A = \emptyset$ {\bf or} $B = \emptyset$ {\bf continue}
              \State $POs[w]$.append(\textsc{SubDAG}($v$, $A$, $B$))
              \State $POs[w]$.append(\textsc{SubDAG}($v'$, $A$, $B$))
          \EndFor
      \EndFor
      \State $Errors \gets \{\}$ \Comment dictionary of errors
      \ForEach{$w \in W$}
        \ForEach{$b \in W$ s.t. $|b| < |w|$}
            \State $g \gets$ \algreplicate($|w|$, $b$, $POs[b]$, $POs[w]$)
            \State $Errors[b][w] \gets$  \textsc{Error}($w$,$g$)
        \EndFor
      \EndFor

      \State \Return {\bf new} $Recipe(W, POs, Errors)$
      \EndFunction
  \end{algorithmic}
  \caption{Algorithm to compute a recipe based on a set of real workflow instances.}\label{alg:alg}
\end{algorithm}

The pseudo-code for \alg is shown in Algorithm~\ref{alg:alg}.  Lines 2 to
16 are devoted to detecting all POs in $W$. For each $w$ in $W$, the
algorithm visits $w$'s vertices (Lines~5-15). An arbitrary unvisited vertex
$v$ is visited, and another arbitrary unvisited vertex $v'$ is found, if it
exists, that has the same type-hash as $v$ (Lines~6-7). If no such $v'$
exists then the algorithm visits another vertex $v$ (Line~8). Otherwise, it
marks $v'$ as visited (Line~9)  and computes the set of closest common
ancestor and successor vertices for $v$ and $v'$ (Lines~10-11).  The
pseudo-code of the \textsc{ClosestCommonAncestors} and
\textsc{ClosestCommonDescendants} functions is not shown as they are simple
DAG traversals. If $v$ and $v'$ do not have at least one common ancestor
and one common descendant, then the algorithm visits another vertex $v$
(Line~12).  Otherwise, two POs have been found, which are constructed and
appended to the list of POs that occur in $w$ at Lines~13 and~14. The
pseudo-code for function \textsc{SubDAG} is not shown. It takes as input a
vertex in a DAG, a set of ancestors of that vertex, and a set of
descendants of that vertex. It returns a DAG that contains all
paths from all ancestors to all descendants to traverse $v$, but from which
the ancestors and descendants have been removed (along with their outgoing
and incoming edges).

Lines~17~to~23 are devoted to computing a set of
``errors" resulting from using a particular (smaller) base workflow to generate
a larger (synthetic) workflow. The \tool approach consists in replicating
POs in a base workflow to scale up its number of vertices while
retaining a realistic structure. Therefore, when needing to generate a
synthetic workflow at a particular scale, it is necessary to choose a base workflow
as a starting point. To provide some basis for this choice, for each $w \in
W$, the algorithm generates a synthetic workflow with $|w|$ vertices using as a
base each workflow in $W$ with fewer vertices than $w$ (Lines~12-22). The
\algreplicate function replicates POs in a base workflow to generate a
larger synthetic workflow (it is described at the end of this section). The error,
that is the discrepancy between the generated workflow  and $w$, is
quantified via some error metric (the \textsc{Error} function) and recorded
at Line~21 (in our implementation we use the THF metric described
in Section~\ref{sec:eval-structure}). The
way in which these recorded errors are used in our approach is explained in
the description of \alggenerate hereafter.  Finally, at Line~24, the
algorithm returns a recipe, i.e., a data structure that contains the
workflow instances ($W$), the discovered pattern occurrences ($POs$), and
the above errors ($Errors$).

\begin{algorithm}[!ht]
    \begin{algorithmic}[1]
        \Function{\alggenerate}{$rcp$, $n$}
        \State $closest \gets$ $w$ in $rcp.W$ s.t. $||w| - n|$ is minimum
        \State $base \gets$ $w$ in $rcp.W$ t. $rcp.Errors[w,closest]$
        \Statex ~~~~is minimum
        \State $g \gets$ \algreplicate($n$, $base$, $rcp.POs[base]$,
        \Statex ~~~~~~~~~~~~~~~~~~$r.POs[closest]$)
        \State \Return $g$
        \EndFunction
    \end{algorithmic}
    \caption{Algorithm for generating a synthetic workflow with $n$ vertices based on a recipe.}\label{alg:alggenerate}
\end{algorithm}

The pseudo-code for \alggenerate is shown in
Algorithm~\ref{alg:alggenerate}. It takes as input a recipe
($rcp$) and a desired number of vertices $n$. At Line~2, the algorithm
determines the workflow in $W$ that has the numbers of vertices closest to
$n$.  This workflow is called $closest$.  At Line~3, the algorithm finds
the workflow in $W$ that, when used as a base for generating a synthetic
workflow with $|closest|$ vertices, leads to the lowest error. The intent
here is to pick the best base workflow for generating a synthetic workflow
with $n$ vertices. No workflow in $W$ may have exactly $n$ vertices. As a
heuristic, we choose the best base workflow for generating a synthetic
workflow with $|closest|$ vertices, based on the errors computed at Lines~17
to~23 in Algorithm~\ref{alg:alg}. The synthetic workflow
is generated by calling function \algreplicate at Line 4, and returned
at Line 5.

\begin{algorithm}[!ht]
    \begin{algorithmic}[1]
        \Function{\algreplicate}{$n$, $base$, $bPOs$, $cPOs$}
        \State $g \gets$ $base$
        \State $prob \gets \{\}$ \Comment dictionary of probabilities
        \ForEach{$po \in bPOs$}
            \State $nc = |\{p \in cPOs\;|\; TH(p) = TH(po)\}|$
            \State $tc = |\{p \in cPOs \}|$
            \State $nb = |\{p \in bPOs\;|\;TH(p) = TH(po)\}|$
            \State $prob[po] \gets (nc/tc) / nb$
        \EndFor
        \While{$|g| < n$}
            \State $po \gets$ sample from $bPO$ with distribution $prob$
            \State $g \gets$ \textsc{AddPO}($g$, $po$)
        \EndWhile
        \State \Return $g$
        \EndFunction
    \end{algorithmic}
    \caption{Algorithm for replicating POs in a base workflow.}\label{alg:algreplicate}
\end{algorithm}

The pseudo-code for \algreplicate is shown in
Algorithm~\ref{alg:algreplicate}.  It takes as input a desired number of
vertices ($n$), a base workflow ($base$), the list of POs in the base
workflow ($bPOs$), and the list of POs in the workflow whose
number of vertices is the closest to $n$
($cPOs$).   The intent is to replicate POs in the base workflow, picking
which pattern to replicate based on the frequency of POs for that pattern in the
closest workflow.  At Line~2, the algorithm first sets the generated
workflow to be the base workflow.  Lines~4-9 are devoted to
computing a probability distribution. More specifically, for each PO in
$bPOs$, the algorithm computes the probability with which this PO should be
replicated. Given a PO in $bPOs$, $nc$ is the number
of POs for that same pattern in $cPOs$ (Line~5) and $tc$ is the total
number of POs in $cPOs$. Thus, $nc/tc$ is the probability that a PO in
$cPOs$ is for that same pattern.  $nb$ is the number of POs in $bPOs$ for
that same pattern (Line~7).  The probability of picking one of these POs in
$bPO$ for replication is thus computed as $((nc/tc)/nb)$ (Line~8).  Note
that this probability could be zero since $nc$ could be zero.  The
algorithm then iteratively adds one PO from the base graph to the generated
graph (while loop at Lines~10 to~13).  At each iteration, a PO $po$ in
$bPO$ is picked randomly with probability $prob[po]$ (Line~11), and this
pattern is added to $g$ (Line~12). The function \textsc{AddPO} operates as
follows. Given a workflow, $g$, and a to-be-added PO, $po$, for a specific
pattern, it: (i)~randomly picks in $g$ one existing PO for that same
pattern, $po'$; (ii)~adds $po$ to the workflow, connecting its entry, resp.
exit, vertices to the parent, resp. children, vertices of the corresponding
entry, resp. exit,  vertices of $po'$.

The pseudo-code in this section is designed for clarity.  Our actual
implementation, described in the next section, is more efficient and
avoids all unnecessary re-computations (e.g., the
probabilities computed in \alggenerate).

\subsection{Implementation}
\label{sec:implementation}

We have implemented our approach in a Python package called \texttt{wfchef}.
Specifically, this package defines a \texttt{Recipe} class. The constructor
for that class takes as input a list of workflow instances and implements
algorithm \alg. The workflow instances are provided as files in the WfCommons
JSON format~\footnote{\url{https://github.com/wfcommons/workflow-schema}}.
The class has a public method \texttt{duplicate} that
implements the \alggenerate algorithm, and a private method
\texttt{duplicate\_nodes} that implements the \algreplicate algorithm. This
Python package is available on
GitHub\footnote{\url{https://github.com/tainagdcoleman/wfchef}}.

\section{Experimental Evaluation}
\label{sec:evaluation}

% \HC{It's probably way too late for this, but often when I read papers that described heuristics, I always wonder what part of the heuristic is really useful. For instance, what if we didn't use the Error stuff to pick the best base DAG, but always
% picked the smallest DAG. How much worse would the results be? Perhaps this is
% easy to do. That way the reviewer gets a sense of what's useful/needed. Otherwise
% one could wonder "what if I pick the base randomly, is that really bad? Perhaps
% you're doing all that fancy stuff when random would be enough...  Food for thought}
% \TC{That would be good. It is possible to do it for THF, but I don't think it is possible to do it for AED. More interesting than the smallest, is the smallest that has all patterns. Journal version maybe?!}
% \HC{Yes, perhaps count the number unique TH of the POs in all workflows, and pick as the base the one that has the largest number. That would be interesting and definitely for a journal paper, in which we could compare 4 approaches: the smallest, random, based on Error as in this work, and based on \# of patterns.}

In this section, we evaluate our approach and compare it to
previously proposed approaches.  In
Section~\ref{sec:methodology}, we describe our experimental methodology. We
evaluate the realism of generated workflows based on their structure,
in Section~\ref{sec:eval-structure}, and based on their simulated
execution, in Section~\ref{sec:eval-comparison}.

\subsection{Methodology}
\label{sec:methodology}

% \HC{Fill in all XXX in the text below}

%\HC{Should we mention other applications at all?}
%\RS{I have added some sentences about that. let me know what you think}
We compare the realism of the synthetic workflow instances generated by
\tool generators to that of instances generated with the original workflow
generator in~\cite{ferreiradasilva-escience-2014}, which is called {\bf
\generator}, and with the more recent generator proposed
in~\cite{ferreiradasilva2020works}, which is called {\bf \workflowhub}.
Recall that both \generator and \workflowhub are hand-crafted, while
\tool generators and automatically produced.

We consider workflow instances from two scientific applications:
(i)~Epigenomics, a bioinformatics workflow~\cite{juve2013characterizing};
and (ii)~Montage, an astronomy workflow~\cite{rynge2013adass}.  We choose
these two applications because they are well-known and used in production.
But also, \generator and \workflowhub can generate synthetic workflow
instances for both of these applications, which makes it possible to
compare our approach to previous work. Note that both these previously
proposed generators support several scientific workflow applications.
However, the only ones they have in common are Epigenomics and Montage.
Two other applications are supported by \generator and also by
\workflowhub.  Unfortunately, \generator generates synthetic workflow
structures that are no longer valid with respect to the current versions of
these applications. Therefore, they are incorrect and not comparable to
real workflow instances or to synthetic workflows generated by
\workflowhub.

Our ground truth consists of real Montage and Epigenomics workflow
instances.  These instances are publicly available on the WorkflowHub
repository~\cite{ferreiradasilva2020works}. They were obtained based on logs of application
executions with the Pegasus Workflow Management
System~\cite{deelman-fgcs-2015} on the Chameleon academic cloud
testbed~\cite{keahey2020lessons}. Specifically, we consider
14 Montage workflow instances with between 105 and 9807 tasks, and
25 Epigenomics workflow instances with between 75 and 1697 tasks.

We generate synthetic workflow instances with the same number of tasks as
real workflow instances, so as to compare synthetic instances to real
instances.  Both \generator and \workflowhub encode application-specific
knowledge to produce synthetic workflow instances for any desired number of
tasks, $n$.  Instead, \tool generators rely on training data, i.e., real
workflow instances. We use a simple ``training and testing" approach. That
is, for generating a synthetic workflow instance with $n$ tasks, we invoke
\alg with all real workflow instances with $<n$ tasks. For instance, say we
want to use \tool to generate an Epigenomics workflow with 127 tasks. We have
real Epigenomics instances for 75, 121, and 127 tasks. We invoke \alg with the
75- and 121-tasks instances to generate the recipe. We then invoke
\alggenerate, passing to it this receipt and asking it to generate a
127-tasks instance.

\subsection{Evaluating the Realism of Synthetic Workflow Structures}
\label{sec:eval-structure}

% \HC{Fill in all XXX in the text below}

We use two graph metrics to quantify the realism of generated workflows, as described
hereafter.

\medskip
\noindent \textbf{\emph{Approximate Edit Distance (AED) --}} Given a real
workflow instance $w$ and a synthetic workflow instance $w'$,
the AED metric is computed as the approximate
number of edits (vertex removal, vertex addition, edge removal, and edge
addition) necessary so that $w = w'$, divided by $|w|$. Lower
values include a higher similarity between $w$ and $w'$. We compute this
metric via the \texttt{{optimize\_graph\_edit\_distance}} method from the Python's
NetworkX package.  Note that NetworkX also provides a method to compute
an exact edit distance, but its complexity is prohibitive for the size of
the workflow instances we consider. Even though the AED metric can be
computed much faster, because it is approximate, we were able to compute it only for workflow instances
with 865 or fewer tasks for Epigenomics and 750 or fewer tasks for Montage. This is
because or RAM footprint issues (despite using a
dedicated host with 192~GiB of RAM).

\begin{figure*}[!ht]
  \centering
  \includegraphics[width=0.8\linewidth]{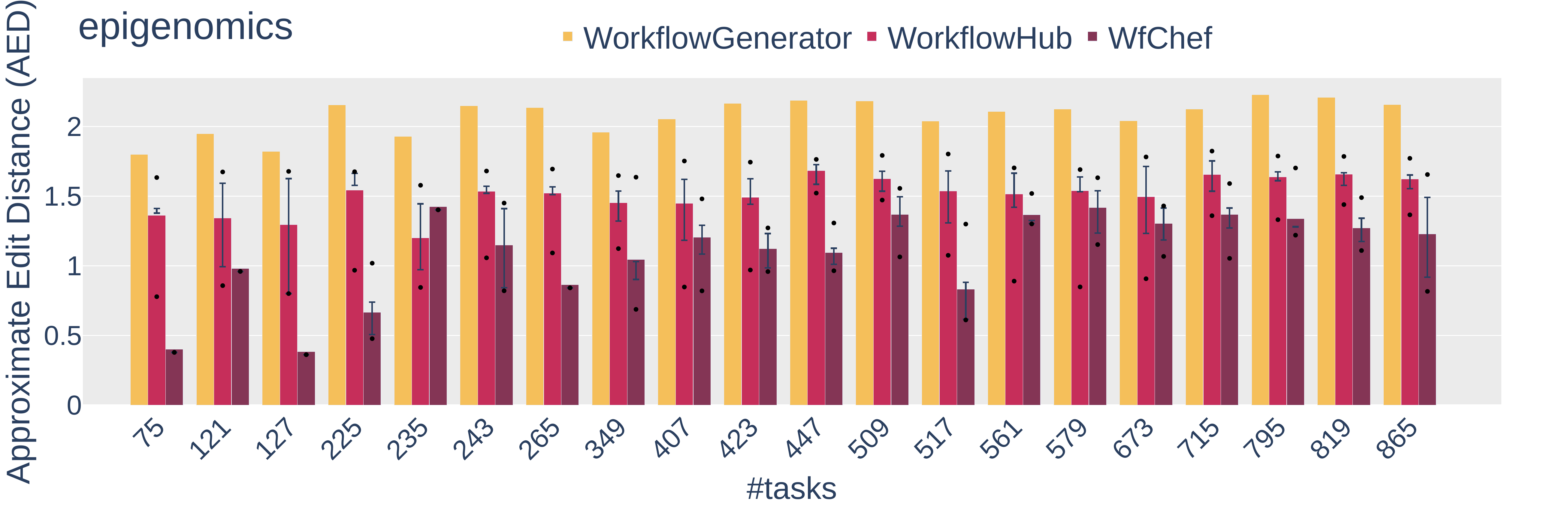}
  \\ \vspace{20pt}
  \includegraphics[width=0.8\linewidth]{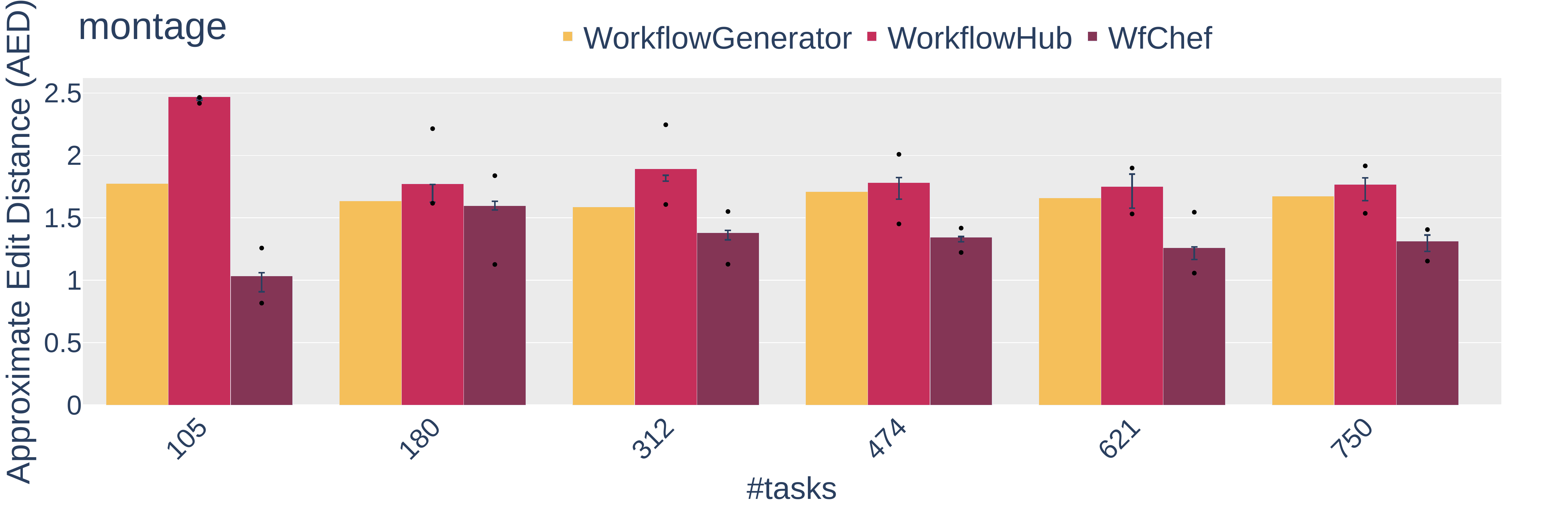}
  \caption{AED for Epigenomics (\emph{top}) and
           Montage (\emph{bottom}) workflows instances.
           Bar heights are average values. Error bars show the
range between the third quartile (Q3)
and the first quartile (Q1), and minimum and maximum values as black dots.
           }
  \label{fig:edit-distance}
\end{figure*}

Figure~\ref{fig:edit-distance} shows AED results for Epigenomics (top) and
Montage (bottom) workflow instances, for \tool, \generator, and
\workflowhub.  \workflowhub and \tool use randomization in their
heuristics. Therefore, for each number tasks we generated 10 sample
synthetic workflow with each tool. The heights of the bars in
Figure~\ref{fig:edit-distance} correspond to average AED values, and we
show error bars that represent the range between the third quartile (Q3)
and the first quartile (Q1), in which  50 percent of the results lie. Error
bars also show minimum and maximum values.  Note that error bars, minimum,
and maximum values are not shown for \generator as it generates synthetic
workflow structures deterministically.

The key observation from Figure~\ref{fig:edit-distance} is that
in most cases \tool leads to lower average AED values than its competitors.
For Epigenomics, \generator leads to
the worst results for all workflow sizes, being significantly outperformed by
\workflowhub.  \workflowhub is itself outperformed by \tool for all
workflow sizes. On average over all Epigenomics instances, \generator,
\workflowhub, and \tool lead to an AED of 2.039, 1.473, and 1.086,
respectively.  For Montage workflows (Figure~\ref{fig:edit-distance}-\emph{bottom}),
\generator outperforms \workflowhub
for all workflow instances, and both are outperformed by \tool. On average over
all Montage instances, \generator, \workflowhub, and \tool lead to an AED of
1.694, 2.216, and 1.111, respectively.

The good results obtained by \tool are due to it being able to generated
instances that are closer in size and that are more faithful to real
workflow instances.  Note that the AED metric values are quite high
overall, often above 1.  Although the synthetic instances may have a
structure that is overall similar to that of the real instances, making the
two workflows absolutely identical requires a large number of edits. For
this reason, hereafter we present results for a second metric.

\medskip
\noindent \textbf{\emph{Type Hash Frequency (THF) --}}
Given a real workflow instance $w$ and a synthetic workflow instance
$w'$, the THF metric is computed as the
Root Mean Square Error (RMSE) of the frequencies of vertex type hashes.
Recall from Definition~\ref{def:th} that the type hash of a vertex encodes
information about a vertex's type but also the types of its ancestors and
successors. Therefore, the more similar the workflow structure
and sub-structures, the lower the THF metric.

% \HC{In Figure~\ref{fig:rms}, change the y axis to say Type Hash Frequency
% (THF)}

\begin{figure*}[!t]
  \centering
  \includegraphics[width=0.8\linewidth]{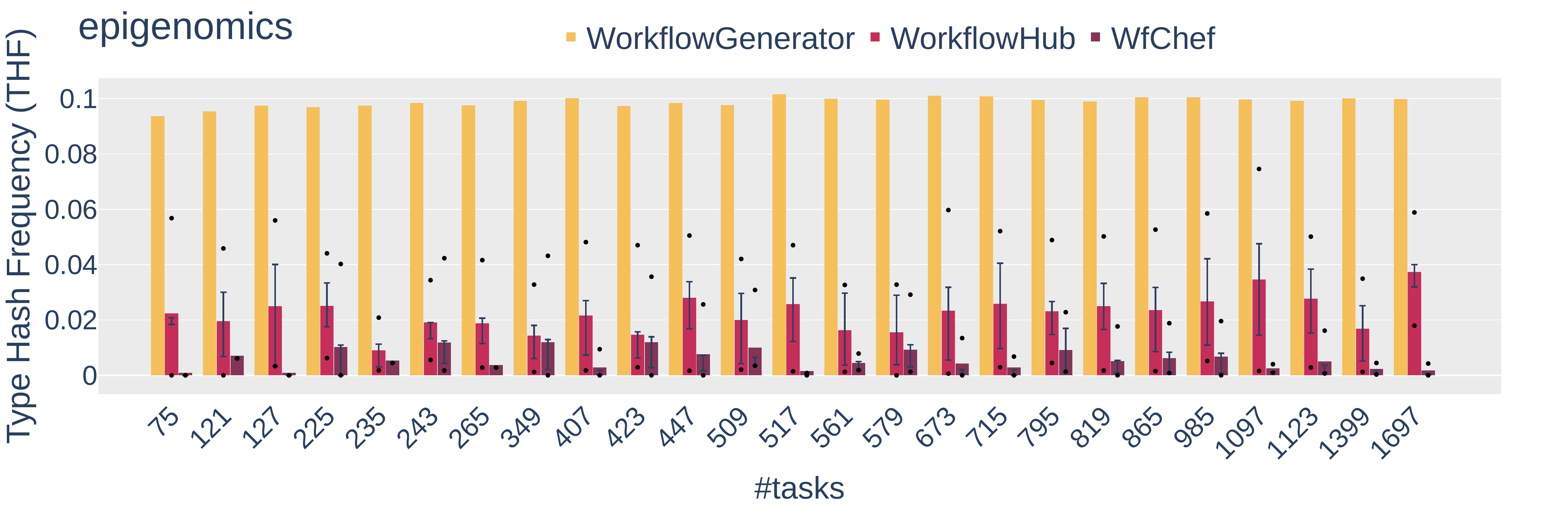}
  \\ \vspace{20pt}
  \includegraphics[width=0.8\linewidth]{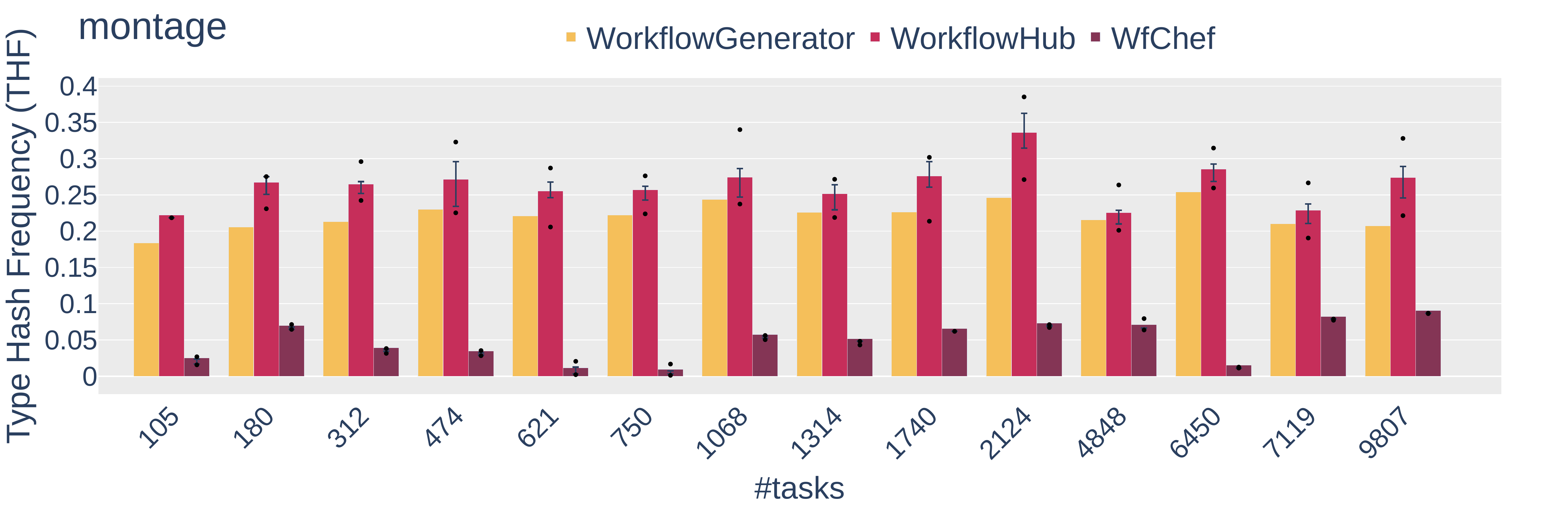}
  \caption{THF for Epigenomics (\emph{top}) and
           Montage (\emph{bottom}) workflows instances.
           Bar heights are average values. Error bars show the
range between the third quartile (Q3)
    and the first quartile (Q1), and minimum and maximum values as black dots.}
  \label{fig:rmse}
\end{figure*}

% \HC{Deal with number of samples for the THF results}

Figure~\ref{fig:rmse} shows THF results for Epigenomics (top) and Montage
(bottom). More results are shown than in Figure~\ref{fig:edit-distance}
since we can evaluate the THF metric for larger workflow instances.
Like in Figure~\ref{fig:edit-distance}, bar heights represent averages
and error bars, minimum, and maximum values are shown for \workflowhub and \tool.

Results are mostly in line with AED results. For Epigenomics, \generator
leads to the worst average results for all workflow sizes. \tool leads to
significantly better results on average than \workflowhub in all but two
cases. For 349- and 423-task workflows, although \tool leads to better
average results, error bars for \tool and \workflowhub have a large amount
of overlap. Note that the length of the error bars for the \tool results
show a fair amount of variation, with short error bars for one workflow
size and significantly longer error bars for the next size up (e.g., going
from 265 tasks to 349 tasks).  This behavior is due to ``jumps" in
structure between workflows of certain scales. In other words, for a given
application, it is common for smaller workflows to contain only a subset of
the patterns that occur in larger workflows.  On average over all
Epigenomics instances, \generator, \workflowhub, and \tool lead to a THF of
0.097, 0.021, and 0.004, respectively. For Montage, \generator leads to
better average results than \workflowhub for all workflow sizes, and \tool leads to
strictly better results than its competitors for all workflow sizes. On average over
all Montage instances, \generator, \workflowhub, and \tool lead to a THF of
0.211, 0.252, and 0.040, respectively.

% but one case (in this case \tool leads to THF that is XXX\% higher than \workflowhub)
%In most cases, \tool leads to results significantly better than \workflowhub
%It is outperformed by \workflowhub in the 7119-task case, and outperformed by both \generator and \workflowhub in the 9807-task case

We conclude that generators produced by \tool generate synthetic
workflow instances with structures that are significantly more realistic than
that of workflows generated by \generator and \workflowhub.

\subsection{Evaluating the Accuracy of Synthetic Workflows}
\label{sec:eval-comparison}

% \HC{Figure 5(b) has weirdly sorted x-axis values}

Synthetic workflow instances are typically used in the literature to drive
simulations of workflow executions. A pragmatic way to evaluate the realism
of synthetic workflow instances is thus to quantify the discrepancy between
their simulated executions to that of their real counterparts, for
executions simulated for the same compute platform using the same
Workflow Management System (WMS).  To do so, we use a
simulator~\cite{wrench-pegasus} of a the state-of-the-art Pegasus
WMS~\cite{deelman-fgcs-2015}. The simulator is built using
WRENCH~\cite{wrenchweb}, a framework for implementing simulators of WMSs
that are accurate and can run scalably on a single computer.
In~\cite{casanova2020fgcs}, it was demonstrated that WRENCH provides high
simulation accuracy for workflow executions using Pegasus.  To ensure
accurate and coherent comparisons, all simulation results in this section
are obtained for the same simulated platform specification as that of the
real-world platforms that was used to obtain the real workflow instances
(based on execution logs): 4 compute nodes each with 48 processors on the
Chameleon testbed~\cite{keahey2020lessons}.

% \HC{We should probably say what platform we are simulating. Rafael said it's the same as the WorkflowHub paper, but I didn't see in there a description of the simulated platform...}

We quantify the discrepancies between the simulated execution of a
synthetic workflow instance and that of a real workflow instance with the
same number of vertices, using two metrics. The first metric is the
absolute relative difference between the simulated makespans (i.e., overall
execution times in seconds).  The second metric is the Root Mean Square
Percentage Error (RMSPE) of workflow task start dates.  The former metric is
simpler (and used often in the literature to quantify simulation error),
but the latter captures more detailed information about the temporal
structure of the simulated executions.

\begin{figure*}[!t]
  \centering
  \includegraphics[width=.9\linewidth]{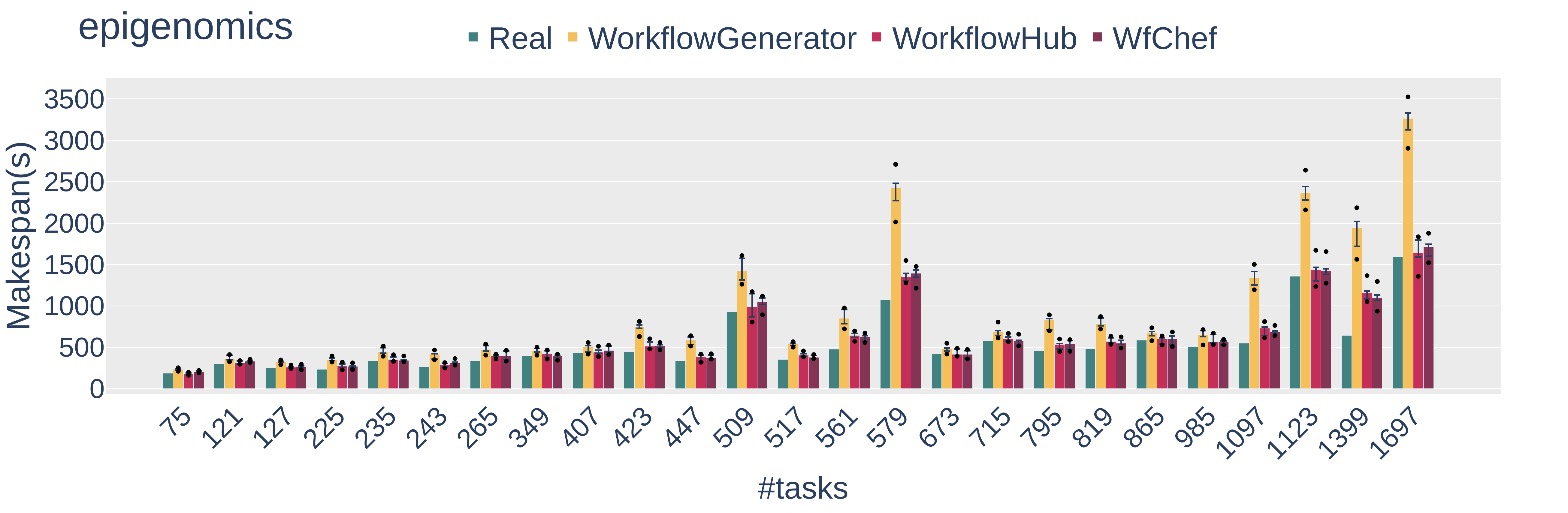}
  \\ \vspace{20pt}
  \includegraphics[width=.9\linewidth]{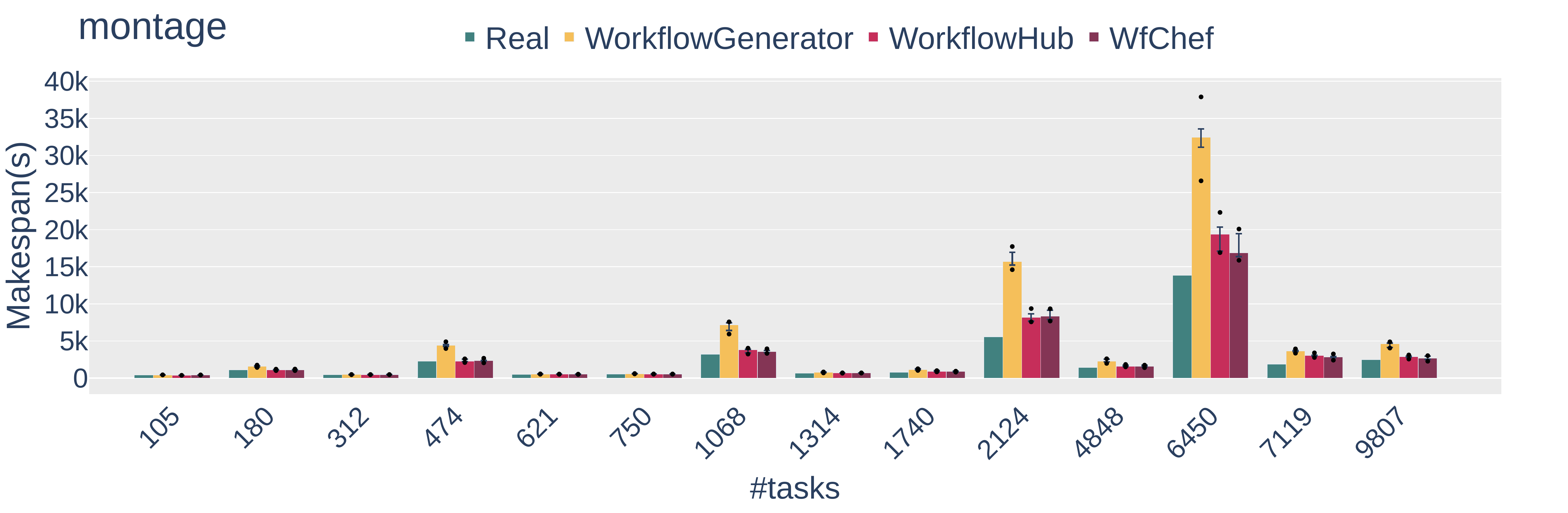}
  \caption{
      Simulated makespan for Epigenomics (\emph{top}) and
           Montage (\emph{bottom}) workflows instances.
           Bar heights are average values. Error bars show the
range between the third quartile (Q3)
and the first quartile (Q1), and minimum and maximum values as black dots.
  }
  \label{fig:scalabillity-makespan}
\end{figure*}

% \HC{Figures in this section: \tool should be the last bar in each group of
% bars, like for the figures in the previous section}
% \TC{I am redoing the edit distance one when I get the data from Rafael, so I will change the order on that one instead of in all the other ones, if that is ok with you}

Figure~\ref{fig:scalabillity-makespan} shows makespans of simulated
executions for real workflow instance and synthetic instances generated by
\generator, \workflowhub, and \tool, for Epigenomics (top) and Montage
(bottom). Note that, unlike for the results in the previous section, error
bars are shown for \generator. Although it generates workflows with
deterministic structure, it samples task characteristics (i.e., task
runtimes, input/output data sizes) from particular random distribution.
Both \workflowhub and \tool do a similar sampling, but from distributions
determined via statistical analysis of real workflow instances.

Overall, we find that the execution of synthetic workflows generated by
\generator leads to the least accurate makespans.  \workflowhub and \tool
lead to better results, with a small advantage for \workflowhub.  On
average over all Epigenomics instances, the average relative differences
between makespans of the real workflow instances and of the synthetic
instances generated by \generator, \workflowhub, and \tool are 75.73\%,
15.21\%, and 15.50\%, respectively. For Montage instances, these averages
are 135.12\%, 32.61\%, and 25.59\%.

\begin{figure*}[!t]
  \centering
  \includegraphics[width=.9\linewidth]{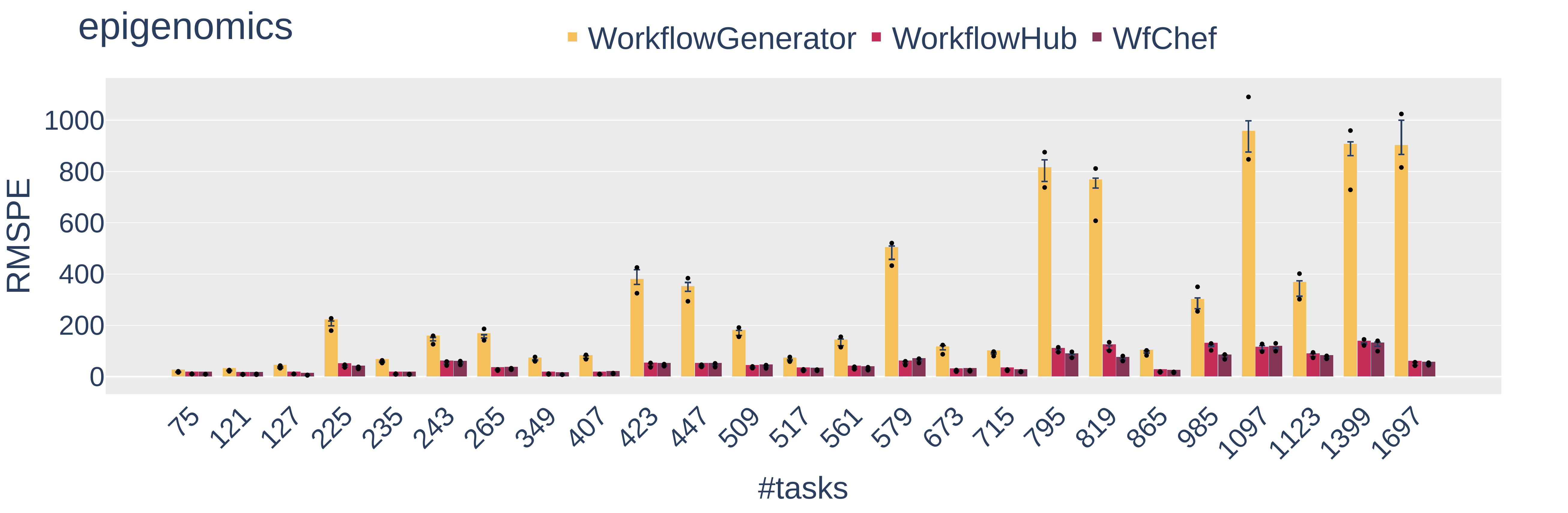}
  \\ \vspace{20pt}
  \includegraphics[width=.9\linewidth]{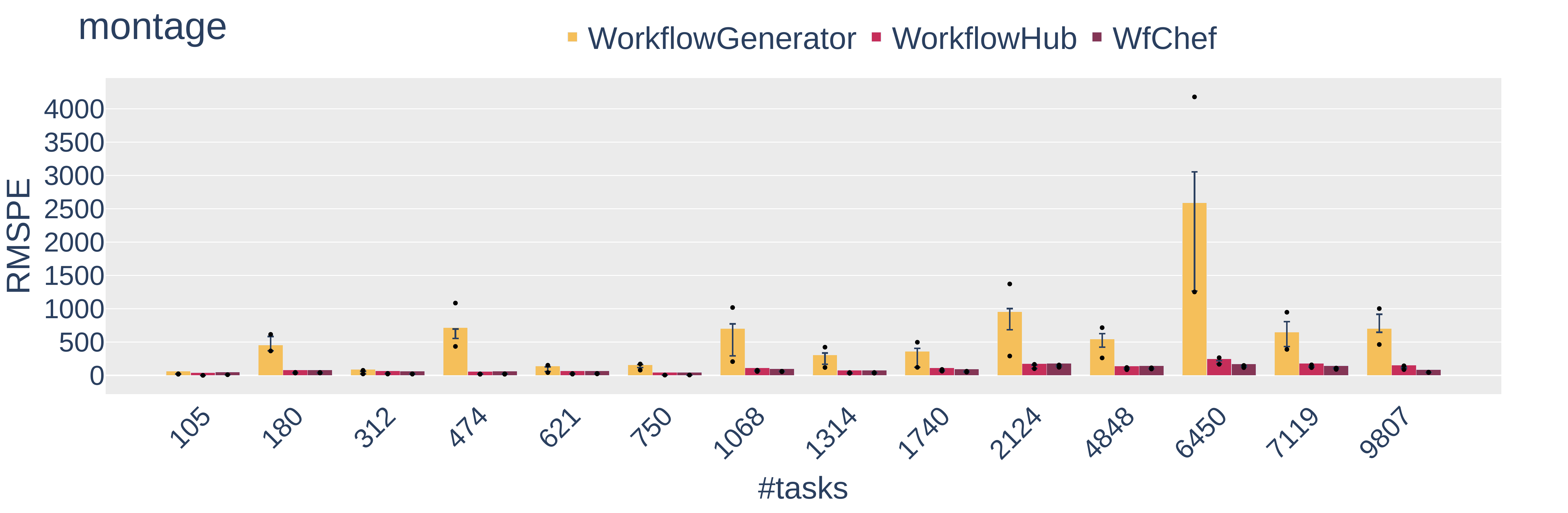}
  \caption{
      RMSPE of simulated task start dates for Epigenomics (\emph{top}) and
           Montage (\emph{bottom}) workflows instances.
           Bar heights are average values. Error bars show the
range between the third quartile (Q3)
and the first quartile (Q1), and minimum and maximum values as black dots.
  }
  \label{fig:scalabillity-rmspe}
\end{figure*}

Figure~\ref{fig:scalabillity-rmspe} shows results for the RMSPE of workflow
task start dates. Here again, we find that the synthetic workflow
instances generated by \generator lead to unrealistic simulated execution.
\workflowhub and \tool lead to more similar results, with a small advantage
to \tool.  On average over all Epigenomics instances, the RMSPE of workflow
task completion dates for synthetic Epigenomics instances generated by
\generator, \workflowhub, and \tool are 294.70\%, 46.08\%, and 40.49\%, respectively.
For Montage instances, these averages are 558.93\%, 64.29\%, and 55.42\%.

We conclude that \tool generators produce synthetic workflow instances
that lead to simulated executions that are drastically more realistic than
that of synthetic workflows generated by \generator.  In fact, it is fair
to say that \generator does not make it possible to obtain realistic
simulation results (which is a concern given its popularity and commonplace
use in the literature). \tool generators lead to results that are similar
but typically more accurate than \workflowhub.  And yet, \tool generators
are automatically generated meaning that, and very much unlike  \generator and
\workflowhub, require zero implementation effort.

\section{Conclusion}
\label{sec:conclusion}
% In this work, we introduced WfChef, a tool to automatically generate realistic synthetic workflow recipes indenpendent of size and application. We proposed a set of algorithms that

% \HC{Drafted this. see what you think}

The availability of synthetic but realistic scientific workflow instances
is crucial for supporting research and development activities in the area
of workflow computing, and in particular for evaluating workflow algorithms
and systems. Although synthetic workflow instance generators have been
developed in previous work, these generators were hand-crafted using expert
knowledge of scientific applications. As a result, their development is
labor-intensive and cannot easily scale to supporting large number of
scientific applications. As an alternative, in this work we have presented
\tool, a tool for automatically generating generators of realistic
synthetic scientific workflow instances. Given a set of real workflow
instances for a particular scientific application, \tool analyzes these
instances in order to discover application-specific patterns. A synthetic
workflow instance with any number of tasks can then be generated by
replicating these patterns in a real workflow instance with fewer tasks.
We have demonstrated that the \tool generators, which require zero software
development efforts, generate more realistic synthetic workflow instances
than the previously available hand-crafted generators. We quantified
workflow instance realism both based on workflow DAG metrics and on
simulated workflow executions.

% \HC{Below I point to the Workflowhub Web site (generators) and that site shows 9 generators for 9 applications. The reviewer may say: why did you compare to only 2 applications when Workflowhub contains 9. I guess our justification is that \generator did not support these....}

A short-term future work direction is to replace the hand-crafted
\workflowhub generators developed in~\cite{ferreiradasilva2020works} and
available on the WorkflowHub web
site\footnote{\url{https://workflowhub.org/generator}} by generators
automatically generated by \tool. Another short-term future direction is to
apply \tool to more scientific workflow applications beyond those supported
by \workflowhub. A longer-term direction is to investigate whether machine
learning techniques can be applied to solve the synthetic workflow
generation problem, to compare these techniques to \tool, and perhaps
evolve \tool accordingly.  Our suspicion, however, is that the amount of
training data necessary for machine learning approaches to be effective
could be prohibitive. By contrast, the \tool algorithms are able to analyze
a few real workflow instances to discover patterns.  In fact, another
interesting research direction is to determine the minimum amount of
training data that still allows \tool to produce realistic synthetic
workflow instances. In the results presented in this work, for the purpose
of evaluating \tool and of comparing it to previously proposed approaches,
we use as training data all available real workflow instances with fewer than the
desired number of workflow tasks. But it may be that using fewer such
instances would still lead to good results.

%% acknowledgments
\vspace{6pt}
\noindent \textbf{Acknowledgments.}
This work is funded by NSF contracts \#1923539 and \#1923621;
and partly funded by NSF contracts \#2016610, and \#2016619.
We also thank the NSF Chameleon Cloud for providing time grants
to access their resources.

% bibliography
\bibliographystyle{IEEEtran}
\bibliography{references}

\balance

\end{document}